\def\selectedoptions{}
\ifx\empty\selectedoptions
  \def\selectedoptions{final}
\fi

\documentclass[
   \selectedoptions
  ]
  {aipproc}

\def\selectedlayoutstyle{6x9}
\layoutstyle\selectedlayoutstyle

\SetInternalRegister\hbadness{8000} 

\newcommand\doingARLO[2][]{%
  \ifx\mmref\undefined #1\else #2\fi
}

\begin{document}

\title 
      [CANGAROO-II and CANGAROO-III]
      {CANGAROO-II and CANGAROO-III}


\author{Masaki Mori}{
  address={Institute for Cosmic Ray Research, University of Tokyo,
  Kashiwa, Chiba 277-8582, Japan\\ E-mail: {\tt morim@icrr.u-tokyo.ac.jp}},
  email={morim@icrr.u-tokyo.ac.jp},
  thanks={}
}

\iftrue
\author{the CANGAROO collaboration}{
  address={Institute for Cosmic Ray Research, Univ.\ of Tokyo; 
      Univ.\ of Adelaide, Ibaraki Univ.;
      Institute of Space and Astronautical Science; 
      National Astronomical Observatory of Japan;
      Tokai Univ.; Tokyo Institute of Technology;
      Kyoto Univ.; STE Laboratory, Nagoya Univ.;
      Yamagata Univ.; Yamanashi Gakuin Univ.; Osaka City Univ.;
      Konan Univ.; Ibaraki Prefectural Univ. Health Sciences;
      Australian National Univ.; Shinshu Univ.},
  email={},
}

\fi

\copyrightyear  {2001}

\begin{abstract}
Preliminary results from CANGAROO-II, a 10~m imaging Cherenkov telescope, 
in Woomera, South Australia are presented. They include the confirmation of 
detections of TeV gamma-ray sources we have reported using a 3.8~m 
telescope, CANGAROO-I. Also the status of the construction of an array 
of four 10~m telescopes, called CANGAROO-III, is reported. 
The first telescope of the array was upgraded from the 7~m telescope and 
the second one is being constructed in this year. 
The full array will be operating in 2004. 
\end{abstract}

\date{\today}

\maketitle

\section{Introduction}


CANGAROO is an acronym for the Collaboration of Australia and Nippon (Japan) 
for a GAmma Ray Observatory in the Outback.
After successful operation of the 3.8m imaging Cherenkov telescope 
(CANGAROO-I) \cite{Har93} for 7 years, which was the first of this
kind in the southern hemisphere,
we constructed a new telescope of 7m diameter (CANGAROO-II) in 1999
\cite{Tan99} next to the 3.8m telescope in Woomera, South Australia
(136$^\circ$47$'$E, 31$^\circ$06$'$E, 160m a.s.l.).
Then the construction of an array of four 10m telescope (CANGAROO-III)
was approved and as a first step the 7m telescope was upgraded
to 10m diameter, which is the first telescope of the CANGAROO-III array 
\cite{Mor00,Mor01,Mor01a,Tan01,Eno01a}.
The major parameters of the CANGAROO telescopes are summarized in Table \ref{7to10m}.

\begin{table}[b]
\begin{tabular}{lccc} \hline
 & 3.8m telescope & 7m telescope & 10m telescope \\ \hline
Focal length & 3.8m & 8m & 8m \\
Number of mirrors & 7 (11m$^2$) & 60 (30m$^2$) \dag & 114 (57m$^2$) \dag \\
Number of PMTs & 256 (3/8") & 512 (1/2") & 552 (1/2") \\
Readout & TDC \& ADC & TDC & TDC \& ADC \\
Point image size (FWHM) &$0.1^\circ$ & 0.15$^\circ$ & 0.20$^\circ$ \\
Operation & 1992--1998 & May 1999--Feb 2000 & Mar 2000-- \\ \hline
\end{tabular}
\caption{Properties of the CANGAROO telescopes. 
 (\dag: CFRP mirrors of 80cm in diameter)}
\label{7to10m}
\end{table}

\begin{figure}
\caption{The 10m imaging Cherenkov telescope in Woomera, South Australia. 
   The huts contain the electronics and the telescope power.}
\includegraphics[height=.4\textheight]{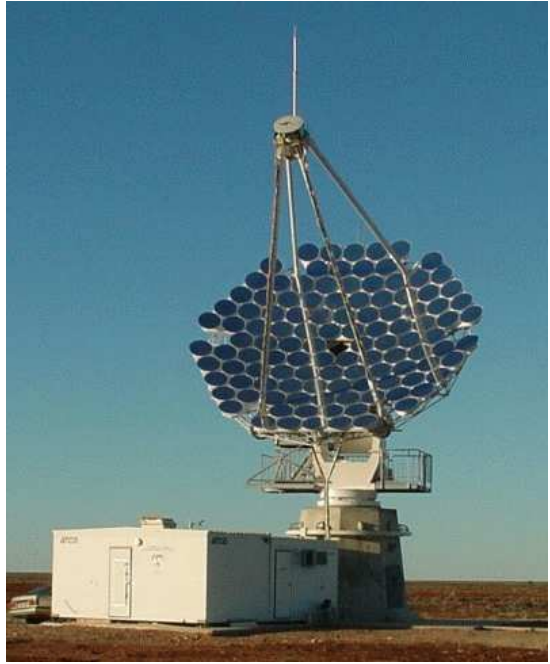}
\end{figure}

\section{Preliminary Results from CANGAROO-II}

Initial performance of the 7m telescope is reported in ref.\cite{Kub00}.
Observations with the 7m and 10m telescope were carried out
in 1999 and 2000, respectively.
The target objects were selected from our list of TeV gamma-ray
sources: Crab, PSR 1706-44, Vela, SN1006, RXJ 1713.7-3946, 
in order to confirm our previous detections with the 3.8m
telescope. Also nearby X-ray selected BL Lacs were observed:
PKS 2005-489, PKS 2155-304 and PKS 0548-322 along with
multiwavelength campaigns. 
Here we give brief description of some preliminary results
obtained so far. 

\paragraph{Crab}

\begin{figure}
\caption{The flux of the Crab obtained from observation with the 7m telescope
based on 43 hours of on- and 40 hours of off-source data. (Preliminary)}
\includegraphics[height=.3\textheight]{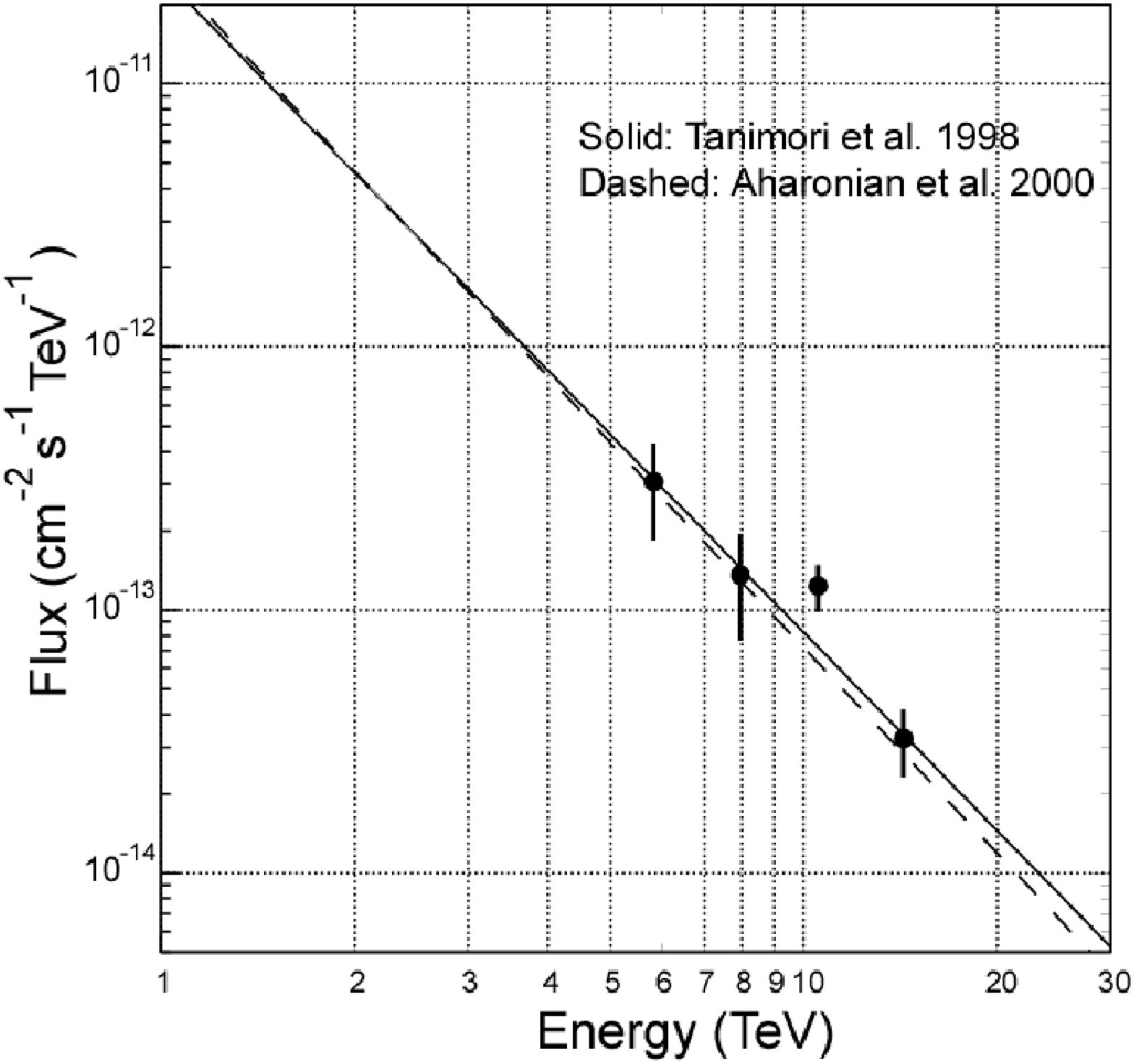}
\label{flux}
\end{figure}

As the standard candle in the TeV gamma-ray astronomy,
we observed the Crab repeatedly, although it is visible only
at large zenith angles ($53^\circ\sim56^\circ$) and thus at 
higher threshold energies ($\sim6$ TeV for 1999 observations)
compared with northern Cherenkov telescopes.
The flux obtained from observation by the 7m telescope
is consitent with our previous report by the 3.8m telescope \cite{Tan98}
(Fig.\ \ref{flux}).

\paragraph{RXJ 1713.7$-$3946 (G347.3$-$0.5)}

\begin{figure}
\caption{The alpha plot of RXJ1713.7$-$3946 data in 2000. Plots are
for on-source data and histograms are for off-source data (11 hours each). 
The excess events near
alpha of zero indicate a gamma-ray signal. (Preliminary)}
\includegraphics[height=.3\textheight]{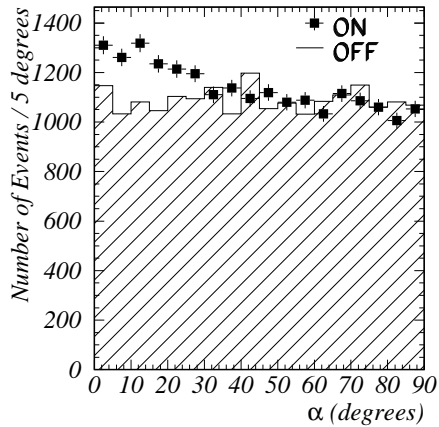}
\label{alpha}
\end{figure}

This is a supernova remnant detected with the 3.8m telescope
\cite{Mur00}.
Figure \ref{alpha} shows the alpha distribution for the data
taken in 2000 with the 10m telescope after the standard imaging 
analysis, showing we have confirmed the detection. 
The threshold energy is $\sim400$ GeV. The peak
near alpha of zero is broader than that for point sources
and may indicate the emission is extended.
The details will be given elsewhere \cite{Eno01}.

\paragraph{PKS 2005$-$489 and PKS 2155$-$304}

Although we spent a lot of time observing these southern nearby BL Lac
objects, we could only set upper limits for these sources.
From observations in 2000 Jul/Aug/Sep, they are 
$6.4\times10^{-12}$ cm$^{-2}$s$^{-1}$ above 450 GeV and
$1.2\times10^{-11}$ cm$^{-2}$s$^{-1}$ above 400 GeV, respectively
 ($2\sigma$ level, preliminary).
The details will be given elsewhere \cite{Nis01}.

\paragraph{Mrk 421}
Following the alert of TeV flaring activities by the HEGRA group, 
we started observation
of Mrk 421 at large zenith angles ($69\sim72^\circ$) in Feb/Mar 2001.
Although the observation time is limited, preliminary analysis shows
a gamma-ray signal at $5\sigma$ level above 9 TeV. 
A detailed analysis is continuing \cite{Oku01}.

\paragraph{Other sources} Results on SN1006 and PSR 1706-44 will be presented 
elsewhere \cite{Har01, Kus01}.

%
%
%

\section{Status of CANGAROO-III}

We will start the construction of the second 10m telescope at the end of 2001.
The full array of four telescopes, set at the corners of a diamond
with sides of about 100m, will be operational in 2004.
The performance as a system of telescopes will be described
elsewhere \cite{Eno01a}.
With experience from the construction and operation of the
CANGAROO-II telescope, we are making following efforts to improve
the sensitivity of the telescopes \cite{Mor01a}.

\paragraph{Reflector}
The reflector design is the same as the first 10m telescope.
The mirrors, made of CFRP \cite{Kaw01}, are light and
have proven to be durable, but they are under further improvement,
especially to obtain better optical quality by
refining the production process.
The mirror attitude adjustment system has been redesigned to
match our needs and save cost.

\paragraph{Telescope control}
Each alt-azimuth telescope is controlled by a PC running Linux with a 
realtime extention (KURT).
A master PC issue directives to each control PC via
network and tracking modes can be flexibly changed.
Clocks are synchronized by NTP software to a GPS receiver.

\paragraph{Camera}
The new design of an imaging camera at the prime focus
is hexagonal shape to minimize the dead space between PMTs.
The total field-of-view is about 4 degrees covered with
427 PMTs of 3/4" diameter \cite{Eno01a}. 
The light guides have been redesigned to maximize photon collection
for the new hexagonal arrangement.
High volages are supplied to PMTs individually.
Each PMT base is included a preamplifier and signals
are transmitted via twisted cables to the electronics
which will be installed at the verandah of the telescope.

\paragraph{Electronics}
The new electronics are all based on the VME specification.
The frontend module amplifies signal and feeds to an ADC, 
discriminates it and feeds to a TDC, an internal scaler and 
a trigger circuit.
The ADC is an improved version of the module used in the
first 10m telescope and includes an internal delay of 150ns
which eliminates a long external delay cable.
A pattern trigger circuit using a Programmable Logic Device 
is under developement to decrease accidental triggers 
due to night-sky background photons.
Details will be given elsewhere \cite{Kub01}.

\paragraph{Monitor}
Cloud monitors detect infrared radiation from clouds making use
of a thermopile module and
supply useful information on data quality \cite{Cla98}.
Weather monitors can record temperature, humidity and wind
speed. These data are read out via serial line connection
and stored for offline analysis.

\section{Summary}

The CANGAROO-II 7/10m telescope has been in operation since 1999
and we have begun to produce preliminary results
which confirm detections made with the
CANGAROO-I 3.8m telescope.
This is the first telescope of an array of four
telescopes, called CANGAROO-III, which will be in
operation in 2004.
The final goal will be an energy threshold of
100 GeV
and an angular resolution of less than 0.1 degee.

\begin{theacknowledgments}
We thank Communication Systems Center, Mitsubishi Electric 
Corporation, and DSC Woomera 
for their assistance in constructing the
telescopes. This project is supported by a Grant-in-Aid for
Scientific Research of Ministry of Education,
Culture, Science, Sports and Technology of Japan, and the
Australian Research Council.
\end{theacknowledgments}


\doingARLO[\bibliographystyle{aipproc}]
          {\ifthenelse{\equal{\AIPcitestyleselect}{num}}
             {\bibliographystyle{arlonum}}
             {\bibliographystyle{arlobib}}
          }
\bibliography{morim}

\hyphenation{Post-Script Sprin-ger}
\begin{thebibliography}{18}
\expandafter\ifx\csname natexlab\endcsname\relax\def\natexlab#1{#1}\fi
\providecommand{\enquote}[1]{``#1''}
\expandafter\ifx\csname url\endcsname\relax
  \def\url#1{\texttt{#1}}\fi
\expandafter\ifx\csname urlprefix\endcsname\relax\def\urlprefix{URL }\fi

\bibitem[{Hara, T. et al.}(1993)]{Har93}
{Hara, T. et al.}, \emph{Nucl.\ Inst.\ Meth.\ Phys.\ Res.}, \textbf{A332}, 300
  (1993).

\bibitem[{Tanimori, T.\ et al.}(1999)]{Tan99}
{Tanimori, T.\ et al.}, \enquote{Construction of New 7m Imaging Air
  \v{C}erenkov Telescope of CANGAROO}, in \emph{Proc.\ 26th ICRC}, edited by
  {Kieda, D.\ et al.}, 5, 1999, p. 203.

\bibitem[{Mori, M.\ et al.}(2000)]{Mor00}
{Mori, M.\ et al.}, \enquote{The CANGAROO-III Project}, in \emph{GeV-TeV Gamma
  Ray Astrophysics Workshop}, edited by {Dingus, B.\ L.\ et al.}, AIP
  Conference Proceedings 515, American Institute of Physics, New York, 2000, p.
  485.

\bibitem[{Mori, M.\ et al.}(2001{\natexlab{a}})]{Mor01}
{Mori, M.\ et al.}, \enquote{Status of the CANGAROO-III Project}, in \emph{High
  Energy Gamma-ray Astronomy}, AIP Conference Proceedings 558, American
  Institute of Physics, New York, 2001{\natexlab{a}}, p. 578.

\bibitem[{Mori, M.\ et al.}(2001{\natexlab{b}})]{Mor01a}
{Mori, M.\ et al.}, \enquote{The CANGAROO-III Project: Status Report}, in
  \emph{Proc.\ 27th ICRC (submitted)}, 2001{\natexlab{b}}.

\bibitem[{Tanimori, T.\ et al.}(2001)]{Tan01}
{Tanimori, T.\ et al.}, \enquote{Recent Status of CANGAROO-III Project}, in
  \emph{High Energy Phenomena in Universe}, edited by T.~T. Vanh, Rencontre de
  Moriond, 2001.

\bibitem[{Enomoto, R.\ et al.}(2001{\natexlab{a}})]{Eno01a}
{Enomoto, R.\ et al.}, \emph{Astropart.\ Phys.\ (in press)}
  (2001{\natexlab{a}}).

\bibitem[{Kubo, H.\ et al.}(2000)]{Kub00}
{Kubo, H.\ et al.}, \enquote{Initial Performance of CANGAROO-II 7m Telescope},
  in \emph{GeV-TeV Gamma Ray Astrophysics Workshop}, edited by {Dingus, B.\ L.\
  et al.}, AIP Conference Proceedings 515, American Institute of Physics, New
  York, 2000, p. 313.

\bibitem[{Tanimori, T.\ et al.}(1998)]{Tan98}
{Tanimori, T.\ et al.}, \emph{Astrophys.\ J.}, \textbf{492}, L33 (1998).

\bibitem[{Muraishi, H.\ et al.}(2000)]{Mur00}
{Muraishi, H.\ et al.}, \emph{Astron.\ Astrophys.}, \textbf{354}, L21 (2000).

\bibitem[{Enomoto, R.\ et al.}(2001{\natexlab{b}})]{Eno01}
{Enomoto, R.\ et al.}, \enquote{Likelihood Analysis of sub-TeV Gamma-rays from
  RXJ1713-39 with CANGAROO-II}, in \emph{Proc.\ 27th ICRC (submitted)},
  2001{\natexlab{b}}.

\bibitem[{Nishijima, K.\ et al.}(2001)]{Nis01}
{Nishijima, K.\ et al.}, \enquote{Very High Energy Gamma-Ray Observation of
  Southern AGNs with CANGAROO-II}, in \emph{Proc.\ 27th ICRC (submitted)},
  2001.

\bibitem[{Okumura, K.\ et al.}(2001)]{Oku01}
{Okumura, K.\ et al.}, \enquote{Search for Gamma-ray Above 10 TeV from
  Markarian 421 in High State with CANGAROO-II Telescope}, in \emph{Proc.\ 27th
  ICRC (submitted)}, 2001.

\bibitem[{Hara, S.\ et al.}(2001)]{Har01}
{Hara, S.\ et al.}, \enquote{Observation of TeV Gamma rays from NE-rim of
  SN1006 with CANGAROO-II 10m Telescope}, in \emph{Proc.\ 27th ICRC
  (submitted)}, 2001.

\bibitem[{Kushida, J.\ et al.}(2001)]{Kus01}
{Kushida, J.\ et al.}, \enquote{Observation of PSR1706-44 with CANGAROO-II
  Telescope}, in \emph{Proc.\ 27th ICRC (submitted)}, 2001.

\bibitem[{Kawachi, A.\ et al.}(2001)]{Kaw01}
{Kawachi, A.\ et al.}, \emph{Astropart.\ Phys.}, \textbf{14}, 261 (2001).

\bibitem[{Kubo, H.\ et al.}(2001)]{Kub01}
{Kubo, H.\ et al.}, \enquote{Development of Data Aquisition System of
  CANGAROO-III Telescope}, in \emph{Proc.\ 27th ICRC (submitted)}, 2001.

\bibitem[{Clay, R.\ W.\ et al.}(1998)]{Cla98}
{Clay, R.\ W.\ et al.}, \emph{Publ.\ Astron.\ Soc.\ Aust.}, \textbf{15}, 332
  (1998).

\end{thebibliography}

\end{document}